# Andreev reflections and tunneling spectroscopy on underdoped $Nd_{1.85}Ce_{0.15}CuO_{4-\delta}$

A. Mourachkine

*Université Libre de Bruxelles, Service de Physique des Solides, CP233, Blvd du Triomphe, B-1050 Brussels, Belgium*

**Abstract.** - Andreev-reflection and tunneling studies have been carried out on a single crystal of underdoped $Nd_{1.85}Ce_{0.15}CuO_{4-\delta}$ (NCCO) by using Ag, Pt-Ir and Nb tips to clarify the symmetry of its order parameter. Surprisingly, we observed in the NCCO, on the one hand, the Josephson current (with Nb tip) which indicates the presence of s-wave order parameter, on the other hand, a zero-bias conductance peak (with Ag tip) which is a manifestation of d-wave order parameter. In addition, we find the presence of two distinct energy scales in the underdoped NCCO: the smallest energy scale is associated with the Andreev gap and *c*-axis tunneling gap, and the second scale corresponds to the maximum in-plane tunneling gap. It seems that the phenomenon of superconductivity in electron- and hole-doped cuprates can be understood within a common scheme.

There is a consensus that the predominant order parameter in hole-doped cuprates has the $d_{x^2-y^2}$ (hereafter, d-wave) symmetry [1]. At the same time, many experiments provide a clear evidence for a significant s-wave component in hole-doped cuprates [2-6]. It is widely believed that electron-doped cuprates are s-wave superconductors (SCs) [7-9]. The main experimental evidence for the s-wave symmetry of the order parameter in $Nd_{1.85}Ce_{0.15}CuO_{4-\delta}$ (NCCO) was obtained in early studies of the London penetration depth $\lambda(T)$ in NCCO single crystals at microwave frequencies [7,10]. However, later it was shown that the exponential dependence of $\lambda(T)$ in NCCO can originate not only from the s-wave order parameter but also from the paramagnetism of $Nd^{3+}$ ions [11]. Early tunneling measurements on NCCO supported the s-wave scenario, however, recent measurements show the presence of a zero-bias conductance peak (ZBCP) [12] which is a manifestation of d-wave order parameter [13]. Recent penetration depth measurements at low temperature are consistent with the presence of the d-wave order parameter in NCCO [14,15]. Thus, the contradiction among different penetration depth and tunneling measurements is obvious.

In hole-doped cuprates, the comparison of Andreev-reflection and tunneling data in the underdoped regime shows the presence of two distinct energy scales: the coherence energy scale (Andreev gap) and single-particle excitation scale (Giaever gap), respectively [16]. The presence of two different energy scales in the underdoped cuprates have been also discussed in the context of other experiments (see references in Ref. [17]). The magnitudes of the two gaps (two order parameters) have different dependencies on hole concentration in $CuO_2$ planes, $p$. The magnitude of the Andreev gap, $\Delta_c$, which scales with $T_c$ as $2\Delta_c/k_B T_c \approx 5.5$ has the parabolic dependence on $p$, while the magnitude of the Giaever gap, $\Delta_p$, increases linearly with the decrease of hole concentration [16].

Tunneling measurements on $Bi_2Sr_2CaCu_2O_{8-x}$ (Bi2212) single crystals show the distribution of gap magnitude in the *ab*-plane [18]. The minimum magnitude of the gap corresponds approximately to the Andreev gap in Bi2212 while the maximum magnitude of the gap corresponds to the Giaever gap [17]. However, in other cuprates, the distribution of tunneling gap has not yet been reported in the literature.

In the present work, we discuss Andreev reflections and tunneling spectroscopy carried out on a single crystal of underdoped NCCO by using Ag, Pt-Ir and Nb tips. Surprisingly, we observed in the NCCO, on the one hand, the Josephson current (with Nb tip) which indicates the presence of s-wave order parameter, on the other hand, a zero-bias conductance peak (ZBCP) (with Ag tip) which is a manifestation of d-wave order parameter. In addition, we find the presence of two distinct energy scales in the underdoped NCCO: the smallest energy scale is associated with the Andreev gap and *c*-axis tunneling gap, and the second scale corresponds to the maximum in-plane tunneling gap. By analyzing the data we find that the phenomenon of SC in NCCO and hole-doped cuprates can be understood within a common scheme.

The $Nd_{1.85}Ce_{0.15}CuO_{4-\delta}$ single crystal used in this study was grown by traveling-solvent floating zone method. Details about the crystal growth conditions are described in Ref. [19]. After the growth, the samples were annealed in an oxygen with a pressure of $10^{-4}$ atmosphere at 1000º C for four days. The single crystals studied here and in Ref. [8] are from the same batch, however, our sample is more underdoped in comparison with the single crystals described in Ref. [8]. The NCCO single crystal having 2.5×2×1.5 mm³ dimensions was cut in a few pieces to perform tunneling measurements. The data obtained on different pieces of the crystal can be considered as obtained in the maternal NCCO single crystal. The $T_c$ value was determined by a SQUID magnetometer yielding $T_c$ = 14.2 ± 0.2 K. The electrical contacts were made by attaching gold wires to a crystal with silver paint. Before mounting, the surfaces of the crystals with orientations along or perpendicular to the *c*-axis were cleaved mechanically in the air.

In this study, we use Ag, Pt-Ir and Nb tips to determine the density of states of NCCO since our attempts to perform break-junction measurements on the NCCO single crystals are failed (it is impossible to break *in situ* a NCCO crystal). Ag and Nb (Pt-Ir) wires were electrochemically (mechanically) sharpened to a point of radius less than 1 μm. Point contacts were formed by pressing a tip against a NCCO single crystal by using a differential screw from the break-junction setup described in Ref. [20]. Measurements have been performed at 4.2 K. The qualifying condition for the contact to be in the Sharvin limit is that its size should be smaller that the mean free path in the electrodes [21]. At 4.2 K, the mean free path has been estimated to be as large as 0.5 μm [22]. The tips are the order of micron-size, but, in cuprates, the actual contact size is usually smaller than the size of the tip [22]. The *I(V)* and *dI/dV(V)* characteristics were determined by using a standard lock-in modulation technique.

The main experimental results of the present study are presented in Figs. 1 - 3. In Fig. 1, the upper spectrum with the Josephson current is obtained in a NCCO/Nb point contact. Since the order parameter in Nb has the s-wave symmetry [7] the presence of the Josephson current in the junction indicates the presence of s-wave order parameter in NCCO. This is in an agreement with early studies carried out on NCCO [7-10]. The Josephson product corresponding to the spectrum shown in Fig. 1 is equal approximately to 200 μV. The lower spectrum in Fig. 1 is obtained in a NCCO/Ag point contact. The ZBCPs shown in Fig. 1 and presented in Ref. [12] have the same shapes. The ZBCP was observed in all hole-doped cuprates studied so far [23]. The ZBCP occurs in general due to the formation of zero energy Andreev bound states at surfaces of d-wave SC [13]. Thus, our result is in a good agreement with the recent penetration depth [14,15] and tunneling measurements [12]. In hole-doped cuprates, there are many pieces of evidence showing the presence of the d-wave [1] and s-wave order parameters [2-6]. It seems that the scenario in electron-doped NCCO is similar to that in hole-doped cuprates: the s-wave and d-wave order parameters are both present in NCCO.

Figures 2(a) and 2(b) show tunneling spectra obtained along and perpendicular to the *c*-axis, respectively. The gap-like features in all *c*-axis spectra have approximately the same value of 3.5 meV. On the other hand, there is the distribution of the in-plane tunneling gap between 3.5 meV and 13 meV. The magnitude of the *c*-axis gap coincides with the minimum in-plane gap. Again, there is the similarity between NCCO and hole-doped cuprates: the distribution of the magnitude of the in-plane gap in both types of cuprates [18]. Ekino and Akimitsu [24] first observed some distribution of tunneling gap in slightly underdoped NCCO ($T_c$ = 17-18 K). Kashiwaya *et al.* [8] first pointed out that the *c*-axis tunneling gap in slightly underdoped NCCO ($T_c$ = 17.5 K) is smaller than the in-plane gap. Other tunneling studies presented in the literature did not report a distribution of the magnitude of the in-plane gap because they have been performed on near optimally doped NCCO samples. The low doping of our sample, instead of being a disadvantage, turned out to be an advantage.

The Andreev-reflection spectrum obtained in a NCCO/Ag point contact is shown in Fig. 3. In the case of an ideal metallic contact between the tip and the SC, the theory predicts that the ratio of the conductances at small and large bias is close to the ideal Andreev value of 2 [21]. In Fig. 3, one can see that this ratio is smaller than 2. Andreev-reflection measurements in hole-doped cuprates show that the ratio is always smaller than 2 in cuprates [25,26]. The magnitude of the Andreev gap in Fig. 3 is in a good agreement with the magnitudes of the *c*-axis gap and minimum in-plane gap. In Fig. 3, the comparison of the magnitudes of the Andreev and Giaever gaps shows unambiguously the presence of two energy scales in the underdoped NCCO. Such scenario is identical to the scenario in underdoped hole-doped cuprates [16].

We discuss now the phase diagrams of NCCO and hole-doped cuprates. Recently, it was shown that the $T_c$ (*p*) phase diagram of hole-doped cuprates is also valid for NCCO [27]. We decided to go further and to compare the two energy scales in NCCO and in hole-doped cuprates. Figure 4 shows the phase diagram of hole-doped cuprates [16]. In Fig. 4, we include also the results of the present study by using empirical relation $T_c/T_{c,\,max}$ = 1 - 82.6(*p* - 0.16)$^2$ between the $T_c$ value and hole concentration [28], and we use $T_{c,\,max}$ = 24 K. The empirical expression is, in fact, valid for hole-doped cuprates, but since the $T_c$ (*p*) phase diagrams of NCCO and hole-doped cuprates are the same [27] we can use it in the case of NCCO. In Fig. 4, one can see that there is an excellent agreement between the phase diagram of hole-doped cuprates and two energy scales in the NCCO. In hole-doped cuprates, the d-wave and s-wave order parameters are associated with the Andreev and Giaever gaps, respectively [17,29]. From Fig. 2, it is difficult to make a conclusion about the symmetries of the two gaps in the NCCO: the spectra which correspond to the two energy scales all have V-shaped gap structures. If the coherence scale in NCCO has the d-wave symmetry and the pairing gap has the s-wave symmetry, then, practically, there is no difference between NCCO and hole-doped cuprates. This raises the question, "Are NCCO and its homologues actually electron-doped, or are they in fact hole-doped cuprates?" [30]. The issue of the s- and d-wave order parameters in NCCO and its correspondence to the Andreev and Giaever gaps awaits future investigations.

In summary, we discussed Andreev reflections and tunneling spectroscopy carried out on a single crystal of underdoped $Nd_{1.85}Ce_{0.15}CuO_{4-\delta}$ by using Ag, Pt-Ir and Nb tips. We observed in the NCCO, on the one hand, the Josephson current (with Nb tip) which indicates the presence of s-wave order parameter, on the other hand, the zero-bias conductance peak (with Ag tip) which is a manifestation of d-wave order parameter in the NCCO. In addition, we found the presence of two distinct energy scales in the underdoped NCCO: the smallest energy scale is associated with the Andreev gap and $c$-axis tunneling gap, and the second scale corresponds to the maximum in-plane tunneling gap. None of the conclusions based on our measurements does not contradict to any previous study carried out on NCCO. On the contrary, the present results reconcile previous data. In addition, we find a new result, namely, that the phenomenon of SC in NCCO and hole-doped cuprates can be understood within a common scheme. It seems that the phase diagram of hole-doped cuprates with the two energy scales is also valid for NCCO and, probably, for all electron-doped cuprates.

* * *

I thank T. Ito for providing the NCCO single crystal, Y. Z. Zhang for performing SQUID measurements, S. Kashiwaya, R. Prozorov and R. Deltour for discussions. This work is supported by PAI 4/10.

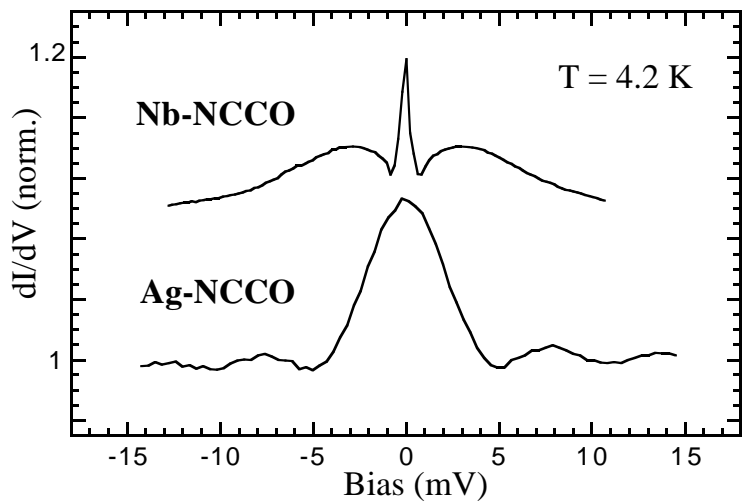

FIG. 1. Normalized d$I$/d$V$ obtained in NCCO/Nb (upper spectrum) and NCCO/Ag (lower spectrum) point contacts. The conductance scale corresponds to the NCCO/Ag spectrum, the NCCO/Nb spectrum is shifted up for clarity.

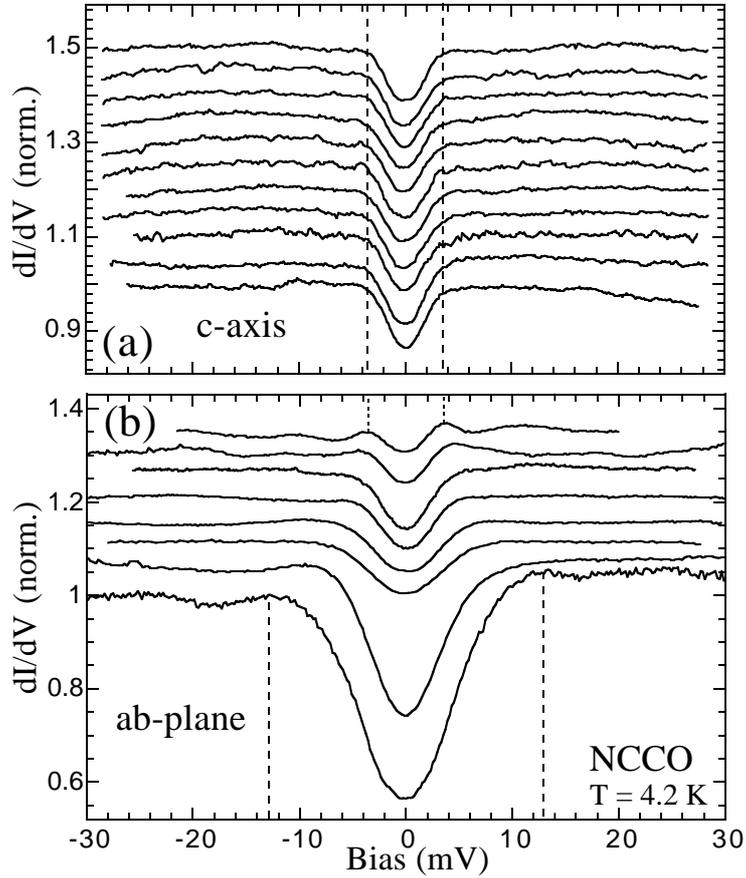

FIG. 2. (a) Spectra obtained in NCCO/Ag point contacts along the *c*-axis. The gap is shown by the dash lines at ± 3.5 meV. (b) Spectra obtained in NCCO/Ag and NCCO/Pt-Ir point contacts perpendicular to the *c*-axis. The dash lines show the minimum and maximum magnitudes of the in-plane gap. In both panels, the conductance scale corresponds to the lower spectrum, the other spectra are offset vertically for clarity.

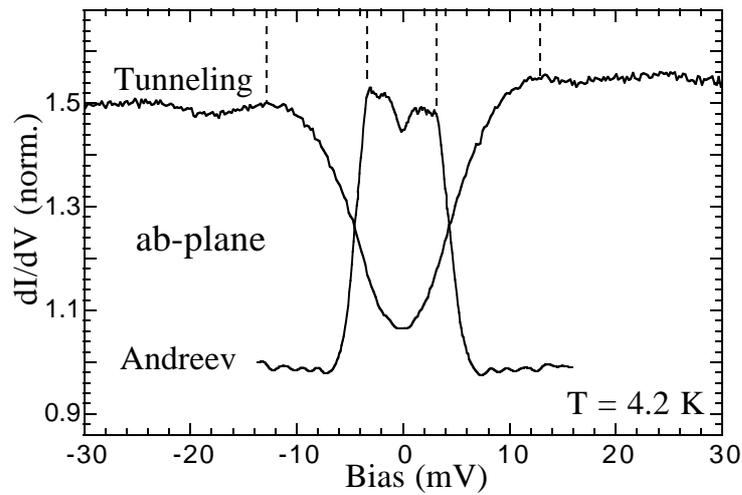

FIG. 3. Andreev-reflection spectrum obtained in a NCCO/Ag point contact and the tunneling spectrum from Fig. 2(b). The conductance scale corresponds to the Andreev-reflection spectrum, the tunneling spectrum is shifted up for clarity. The dash lines show two energy scales.

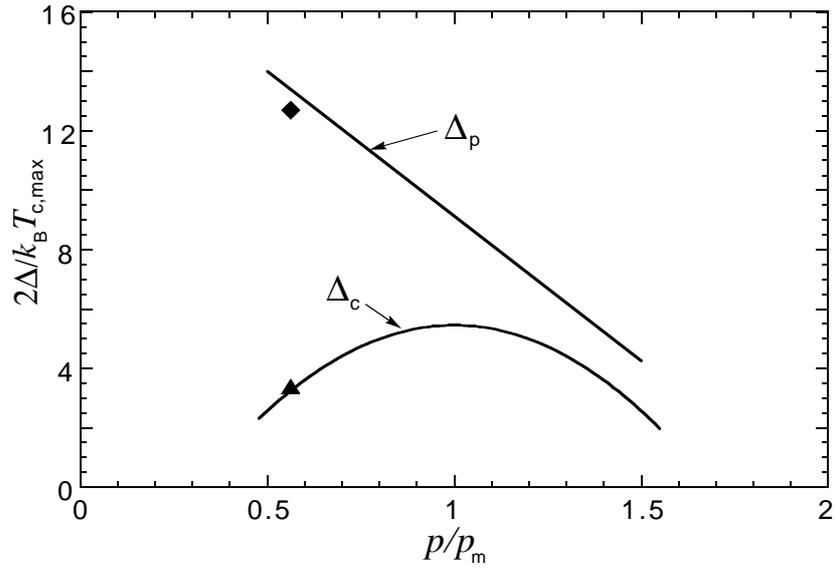

FIG. 4. Phase diagram of hole-doped cuprates at low temperature: $\Delta_c$ is the coherence energy scale, and $\Delta_p$ is the single-particle excitation gap. The $p_m$ is a hole concentration with the maximum $T_c$ (from Ref. 16). The triangle and diamond correspond to the two energy scales in underdoped NCCO, shown in Figs. 3 and 2.